\documentclass[11pt,toc]{JHEP}
\usepackage{cite}
\usepackage{epsfig}
\input epsf
\catcode `@=11 \@addtoreset{equation}{section} \catcode `@=12
\def\be{\begin{equation}}
\def\ee{\end{equation}}
\def\ba{\begin{array}{l}}
\def\ea{\end{array}}
\def\bea{\begin{eqnarray}}
\def\eea{\end{eqnarray}}

\def\nn{\nonumber\\}

\def\nu{\noindent\underbar}

\def\eq#1{(\ref{#1})}

\def\ars{\cite{Alekseev:2000fd}}
\def\myers{\cite{Myers:1999ps}}
\def\bds{\cite{Bachas:2000ik}}

\title{\Large Matrix dynamics of fuzzy spheres}
\author{by
Dileep P. Jatkar$^a$, Gautam Mandal$^b$,
Spenta R. Wadia$^b$ and K.P. Yogendran$^b$\\
$^a$ {\sl Harish-Chandra Research Institute,\\
Chhatnag Road, Jhusi, Allahabad 211 019, India}\\

\vspace*{1ex}

$^b$ {\sl Department of Theoretical Physics,
 Tata Institute of Fundamental Research,\\
Homi Bhabha Road, Mumbai 400 005, INDIA. }\\
e-mail: \email{dileep@mri.ernet.in,
mandal,wadia,kpy@theory.tifr.res.in}}

\preprint{\hepth{0110172}\\ HRI-P-011001 \\TIFR/TH/01-20}

\abstract{We study the dynamics of fuzzy two-spheres in a matrix model
which represents string theory in the presence of RR flux.  We analyze
the stability of the known static solutions of such a theory which
contain commuting matrices and SU(2) representations. We find that the
irreducible as well as the reducible representations are stable.
Since the latter are of higher energy, this stability poses a
puzzle. We resolve this puzzle by noting that the reducible
represenations have marginal deformations corresponding to
non-spherical deformations. We obtain new static solutions by turning
on these marginal deformations. These solutions now have instability
or tachyonic directions. We discuss condensation of these tachyons
which corresponds to classical trajectories interpolating from
multiple, small fuzzy spheres to a single, large sphere. We briefly
discuss spatially independent configurations of a D3/D5 system
described by the same matrix model which now possesses a supergravity
dual.}


\begin{document}

\section{Introduction}

\vspace{-2ex}

Noncommutative spheres have made their appearance in string theory for
a variety of reasons
{}\cite{Myers:1999ps,Taylor:2000pr,Bachas:2000ik,McGreevy:2000cw,
Polchinski:2000uf,Li:2001bd,Alekseev:2000fd,Pawelczyk:2000ah,Johnson:2001bm,
Trivedi:2000mq, Pawelczyk:2000hy}.
Apart from their intrinsic
interest\cite{Johnson:2001bm,Trivedi:2000mq,Vaidya:2001bt, Chu:2001xi,
Iso:2001mg,Johnson:2001di,Ho:2000qh,Bal:2001cs},
non-commutative spaces could actually be a more realistic description
of space-time at very short length
scales\cite{Taylor:1999qk,Taylor:2001vb}.  One way to argue this would
be that \ars\ in a curved space-time generically the NS B-field is
present (by string equations of motion). The fuzzy geometries that we
will encounter \myers arise because of the presence of RR fluxes
through a three-sphere.  One or the other such background field will
generically be present in a curved background.  Indeed,
non-commutativity of space-time may be of a more basic nature, as
indicated by its appearance in string field
theories\cite{Witten:1986cc}.  Fuzzy spheres have also been
investigated in the context of dynamics of giant
gravitons\cite{McGreevy:2000cw, Li:2001bd,
Youm:2001cg,Kim:2001dp,Mikhailov:2000ya,Das:2001st,Das:2000ab,Das:2001fu,
Millar:2000ib}.

In this paper we will study dynamics of fuzzy spheres, indeed more
generic fuzzy two-branes. We will start with the matrix model
action\cite{Banks:1997vh} with a Chern-Simon term, which arises due to
background constant four form flux\myers . In section 2, we set up the
model and its equations of motion. We then review how SU(2)
representations with various spin, as well as commuting matrices, are
solutions to the equation of motion. The nontrivial SU(2)
representations geometrically correspond to fuzzy spheres. The energy
of the various solutions suggests a picture of cascade, where
reducible representations have higher energy than the irreducible
representation. In section 3, we study quadratic fluctuations around
the solutions discussed in section 2. Contrary to our expectations we
find that all the solutions labelled by different SU(2) spin
representations are all stable, i.e., no quadratic fluctuations have
negative mass squared.  This raises a puzzle regarding our cascade
picture. We also briefly discuss some geometric aspects of the
fluctuation spectrum. The puzzle about unstable solutions is resolved
in an interesting way in section 4, by finding new solutions to the
equations of motion, which corresponds to non-spherical configurations.
\footnote{More details of deformations of fuzzy spheres and
interesting issue of topology change due to non-commutativity will be
discussed in detail elsewhere\cite{MWY}.} 
These solutions correspond
to exactly marginal deformations of the original solutions. 
A subset of these solutions has earlier been discussed and
analyzed in \cite{Hashimoto:2001xy}.
We study quadratic fluctuations around these new solutions and find
tachyonic instability.  This is an indication that there exists
at least one solution with lower energy. In the second part of Section
4 we discuss other new and interesting features of quadratic
fluctuations around these new solutions. We show that in certain
cases, due to deformation by a marginal parameter, number of flat
directions increase. This jump in the dimensionality of moduli space
is reminiscent of emergence of new marginal operators in $c=1$
conformal field theory at self dual radius. We also discuss patterns
of symmetry breaking in this section. In section 5, we relate our
method with other approaches. We first compare our results with the
spherical D2 brane of Bachas, Douglas and
Schweigert\cite{Bachas:2000ik}. We also discuss Myers' dual D2
brane\myers. In section 6, we present the energy landscape more
quantitatively and discuss dynamical evolution of multiple fuzzy
spheres into a single big fuzzy sphere.  In Section 7 we discuss
effects of turning on a mass term in our matrix model. In section 7,
we discuss relevance of our results to the dual SUGRA
solution\cite{Polchinski:2000uf} and also to F1-NS5 system in type IIA
string theory.

\section{The model}\label{two}

The matrix model we will be concerned with is described by the action:
\be
S= T_0 \ Tr \Big( \frac12
\dot{X_i}^2 + \frac14  [X_i,X_j][X_i,X_j] -
\frac{i}3  \kappa \ \epsilon_{ijk} X_i[X_j,X_k] \Big)
\label{action}
\ee
where $X_i, i=1,2,3$ are $N \times N$ matrices and
$T_0=\sqrt{2\pi}/g_s$ is
the zero-brane tension. Throughout this paper, we use units such that
$2\pi\alpha'=2\pi l_{s}^{2}=1$. This action must be supplemented by a
Gauss law condition, arising from the $A_0$ equation of motion.

This model arises in several contexts. We will mention two cases
(see Section 5 and Section 8 for more details of
these two and other cases):

(a) Myers \myers\ discusses \eq{action} to describe D0 brane quantum
mechanics (in type IIA theory) in the presence of a constant non-zero
vev of the 4-form flux $F^{(4)}_{tijk}=-2\kappa\epsilon_{ijk}$. Note
that $\kappa$ has dimensions $L^{-1}$.  The last term (the Chern-Simon
term) in the action produces interaction between D0 branes due to the
4-form flux. This interaction is demanded by consistency of the
D-brane action with the T-duality symmetry of the string theory.

(b) Alekseev et al. \ars\ derived this \eq{action} as an effective
action that reproduces the correlation functions of an SU(2) level $k$
WZW BCFT, representing $S^2$ branes wrapping on an $S^3$ (with radius
given by $k$).

The approaches (a) and (b) are, in fact, connected. E.g., \bds\ shows
that the WZW BCFT can be accurately described at large $k$ in terms a
DBI action for D2-branes in the presence a two-form RR flux and a
gauge field background on the brane. These D2-branes are spherical and
are embedded in $S^3$; as one takes the radius of the 3-sphere to
infinity, the two background fluxes mentioned above exactly match with
\myers\ description of those fluxes (in the D2-description).

We will use this model as means to study the dynamics of
non-commutative 2-branes, notably spherical branes.

In the rest of this section we will review the static
solutions of this action discussed in \myers\ and \ars.
The static equations of motion are 
\be
[X_j, \bigg( [X_i,X_j] - i\kappa\  \epsilon_{ijk} X_k \bigg) ]=0.
\label{eom}
\ee
This equation of motion admits an obvious solution  
\be
[X_i,X_j]=0,
\label{comm}
\ee 
which represents $N$ D0 branes (with the $x^i$ coordinates given by
the diagonal elements of the matrices $X_i$). In the absence of the
Chern-Simon term it is well known that this configuration is a lowest
energy configuration satisfying the BPS condition. However, as pointed
out in \myers, the Chern-Simon term modifies the situation
radically. In particular, commuting matrices are no longer lowest
energy configurations. In fact, among the available set of extrema,
this configuration has the highest energy.

Besides, it  admits the following static solution which satisfies:
\be
[X_i, X_j] = i\kappa \ \epsilon_{ijk} X_k.
\label{su2}
\ee
Clearly any matrix representation of SU(2) will satisfy this equation of
motion. It is easy to explicitly write down such $X_i$'s
\be
X_i = \kappa J_i
\label{irrep}
\ee
where $J_1, J_2, J_3$ define, say, the $N$ dimensional irreducible 
representation of su(2).
It is well-known that \eq{irrep} define a fuzzy $S^2$, of
radius $r$, where
\be
X_i X_i = \frac{R^2}{4\pi^2 l^{4}_s}{\bf 1}, \; 
R^2=\kappa^2 j(j+1), \;
2j +1 \equiv N
\label{radius}
\ee
Besides \eq{comm} and \eq{irrep}, \myers, \ars\ also specify reducible
solutions. That is, $X_i$ can be a direct sum of several irreducible
representations of SU(2). Such a configuration also solves the
equation of motion.
\be
X_i = \kappa \oplus_{r=1}^s J_i^{(r)}
\label{rep}
\ee where 
\be 
\sum_{r=1}^s (2 j_r + 1)= N 
\ee 
It is clear from
eq. \eq{radius} that this representation corresponds to $s$ fuzzy spheres
with radii
\be
R_r^2 = \kappa^2 j_r(j_r+1)
\ee
In \eq{rep}, the irreducible representations $j_r$ can also include
the trivial representation $J_i=0$. It is, therefore, obvious that
\eq{irrep} as well as \eq{comm} are special cases of \eq{rep}.

It is important to consider the energy of these static solutions which
is given by the static hamiltonian, (in units of $T_0$, 
the D0-brane tension)
\be 
V=  Tr (-\frac{1}{4}
[X_i,X_j][X_i,X_j] + \frac{i}3 \kappa \ \epsilon_{ijk} X_i[X_j,X_k] ).
\label{ham}
\ee
The energy E of \eq{rep} is given by 
\be
E = - T_0 \kappa^4 \ \frac{1}{6} \sum_{r=1}^{s} j_r (j_r+1)( 2 j_r +1).
\ee
The trivial representation $j_r=0, \, \forall r$ corresponds to the
commuting set of matrices and the energy of this configuration is
zero. It is worthwhile to mention here that we are measuring the
energy of these configuration with respect to the energy of $N$ times
the single D0 brane mass (tension).  Clearly, zero energy for the
commuting matrices is the reflection of the no force condition between
BPS configurations. Nontrivial SU(2) configurations, however, have
negative energy. This means these configurations are more stable than
the trivial configuration. The lowest energy configuration is the spin
$j= (N-1)/2$ dimensional irreducible representation.

%\underbar{Example}:
It is easy to illustrate this fact by taking simple low dimensional
matrix examples. Let us consider an example of $4 \times 4$
matrices. In this case there are five distinct solutions to the
equations of motion. The static solutions, their sizes, i.e., the
radii of the fuzzy spheres and energies are summarized in the table 1
below.

\vspace{3ex}

\centerline{Table 1}

\vspace{2ex}

\begin{tabular}{ | l | l | l | }
\hline
Solution  & Radii of fuzzy spheres  
& Energy \\
	& (in units of $\kappa/2\pi$)&(in units of $T_0 \kappa^4$) \\
\hline
(1) Commuting  & (0,0,0,0)  & 0 \\
(2) spin $\frac{1}{2} \oplus 0 \oplus 0$ & $(\sqrt 3/2,0,0)$ & 
$-\frac{1}{4}$ \\
(3) spin $\frac{1}{2} \oplus \frac{1}{2}$ & $(\sqrt 3/2,\sqrt 3/2 )$
& $-1/2$ \\
(4) spin $1 \oplus 0$ & $(\sqrt 2, 0)$& $-1$ \\
(5) spin $\frac{3}{2}$ & $\sqrt 15/2$ & $-5/2$ \\
\hline
\end{tabular}

\vspace{3ex}

\subsection{Cascade}\label{cascade}

As anticipated from the energy formula, the commuting matrices have
highest energy and the irreducible representation has lowest energy.
The intermediate solutions (2,3,4 in the above table) have energies in
between these two extremes. These energy values suggest the picture of
a ``cascade'', i.e., all the configurations except the irreducible
representation should presumably have instabilities or ``tachyonic''
directions which should lead to their decay into the most stable
configuration, viz. the irreducible representation. Geometrically this
means the tachyonic instability would set off the process of smaller
spheres ``fusing'' into larger spheres which ultimately would turn
into the largest sphere to achieve minimum energy.

\section{Analysis of instabilities}

Let us now look at the instabilities of these solutions.
The quadratic fluctuation of $H$ around a general static
solution has the form
\bea
\delta^2 H &=&
 \frac{\Pi_{i}^2}{2T_0} +
T_0\ {\rm Tr} \Big( -[\delta X_i, X_j] [\delta X_i, X_j]
+ [\delta X_i, X_i] [\delta X_j, X_j] \Big) -
\nonumber\\
&& 2T_0\ {\rm Tr} \Big( \big( [X_i, X_j] - i \kappa \epsilon_{ijk} X_k
\big) [\delta X_i, \delta X_j] \Big)
\label{fluct}\\
&=& \frac12 [\pi_{ia}^2 +
y_{ia} M_{ij,ab} y_{ib}]
\label{def-m}
\eea
In the above we have parametrized the fluctuations $ \delta X_i =
\frac1{\sqrt{T_0}} \sum_a y_{ia} \lambda_a, i=1,2,3;
a=1,2,\ldots\\ ,n^2-1$.  $\Pi_i$ are momenta conjugate to $\delta X_i$
while $\pi_{ia}$ are momenta conjugate to $y_{ia}$.

In order to understand the energy landscape of our model (cf. Sec
\ref{cascade}), it is important first to understand the neighbourhood
of the critical points of the energy function.  We will therefore be
interested in The eigenvalues $\kappa^2 \omega^2_{ia}$ of the
quadratic fluctuation matrix $M_{ij,ab}$ (these are calculated for
various $n$ in Appendix A).

We will find below that for all the critical points \eq{rep} and
\eq{irrep} described so far, $\omega^2\ge 0$, hence there are no
instabilities. This leads to a
puzzle. We will describe first the calculation of eigenvalues
and then return to the puzzle.

\subsection{Spectrum of fluctuations}\label{unpert}

For the solutions  \eq{rep} or \eq{irrep}, the second
line of \eq{fluct} vanishes by virtue of equation of motion \eq{su2}.
Using, further, the gauge fixing
condition, $[X_i,\delta X_i]=0$, we get
\bea \delta^2 H
&&= \frac{\Pi_{i}^2}{2T_0} - \frac{T_0}{2}[X_i,\delta X_j]^2
\label{fluct1}
\\ &&= \frac12 [\pi_{ia}^2 +
y_{ia} M^{(0)}_{ij,ab} y_{ib}] 
\label{def-m0}
\eea
The tables in Appendix A  list eigenvalues
in which marginal deformations $Y$, discussed in the next
section, are turned on; in order to find the eigenvalues
without them we have to turn them off. Let us mention,
for example,  the eigenvalue table  for $3 \times
3$ matrices. We reproduce here the fluctuations around the spin
$1/2 \oplus 0 $ and spin $1$ solutions here.
Absence of the $Y$-deformations means that we have put $\vec c=0$:

\vspace{3ex}
\begin{center}
Table 2: $N=3$\\
\noindent{\small Numbers in square brackets refer to physical zero
modes}\\
\vspace{2ex}
\begin{tabular}{||l|l|l||}\hline
Solution & $ \omega^2 $ & Multiplicity \\ \hline
   &        0 & 10 [3]   \\ \cline{2-3}
     $\frac{1}{2} \oplus 0$          & 2     & 6     \\ \cline{2-3}
               &$(3/4)$     & 8\\
\cline{1-3}
        & 0    & 8 [0]     \\ \cline{2-3}
         $1$        & 2     & 6      \\ \cline{2-3}
               & 6     & 10    \\ \hline
\end{tabular}
\end{center}

\subsubsection{Zero modes}

For the spin $1$ solution, corresponding to the largest sphere for
Mat(3), the 8 zero modes are all of the form $\delta^{(a)} X_i =
[\lambda_a, \bar X_i]$ where $\lambda_a, a=1,..,8$ are $SU(3) $
Gell-Mann matrices.  It is easy to see from the Gauss law $0=[X_i,
\Pi_i]$ that these are gauge rotations.  The notation 8[0] means 8
zero modes, 0 of which are physical.  For the spin $\frac{1}{2} \oplus
0$ the SU(3) rotations produce 7 gauge zero modes, since $[\lambda_8,
\bar X_i]=0$. There are three physical zero modes which are
infinitesimal versions of the three $Y$-deformations.  It is easy to
see the general formula for the degeneracy of exact zero modes is that
there are a total of $\mu_0= n^2-1 + 2 n_a$ zero modes out of which $3
n_a$ are physical (this excludes eigenvalues which become non-zero
after $Y$-deformations).

\subsubsection{Non-zero modes}

Around a spin $j$ irrep \eq{irrep}, like
the spin 1 example above, the eigenvalues $\kappa^2
\omega^2$ of the quadratic fluctuation $M$ in \eq{def-m0}   are 
given by
\be
\omega_l^2 = \sqrt{l(l+1)}, \, l=1,...,2j
\label{eigen}
\ee each with multiplicity $\mu_l= 2(l+1)$. Thus, 
$\omega_1^2=2$ 
appears 6 times and $\omega^2_2=6$ appears 10 times for the spin
1 solution above.  To see this spectrum, note
that  the quadratic fluctuation operator in
\eq{fluct1} is then simply the laplacian $(\hat J_i)^2 \equiv (Ad \bar
X_i)^2$.  The $\hat J_i$ acts here by adjoint action on matrix-type
fluctuations, rather than on column vectors, hence the eigenspaces
split into representations $ l \in j \times j = 0, 1, ..., 2j$ (we
exclude $l=0$ in the above table since we have restricted to traceless
$\delta X_i$).  The degeneracies $\mu_l$ are twice $2l +1$ since there
are two independent fluctuations $\delta X_i, i=1,2$.

By a simple generalization of the above argument, the eigenvalues
around a reducible representation (e.g.  spin $j \oplus j'$) include
(a) the eigenvalue set around an irrep $j$, (b) the eigenvalue set
around an irrep $j'$, and (c) eigenvalues $ \omega_l^2 = l(l+1),
l=|j-j'|,..., j+j'$, with multiplicities $\mu_l = 4 (2l+1)$. In the
spin $\frac{1}{2} \oplus 0$ example above, we have
$\omega^2_{1/2}=3/4$ appearing $\mu_{1/2}=8$ times.

\subsubsection{Geometrical interpretation of eigenvalues}

To put all this more simply, the structure of the fluctuation, say $\delta
X_1$, around a solution spin $j\oplus j'$ is
\be
\delta X_1= 
\left(
\begin{array}{cc}
a_{j\times j}  & c_{j\times j'}
\\
c^\dagger_{j'\times j}  & b_{j'\times j'}
\end{array}
\right)
\ee
Recall that the solution is
\be
\bar X_i= \kappa \left( \begin{array}{cc}{J_i}_{j\times j} & 
0_{j\times j'}\\
			0_{j'\times j} & {J_i}_{j'\times j'}
\end{array} 
\right) 
\ee
The eigenvalues coming from the block $a$, namely $\omega^2_l= l(l+1),
l = 1,...,2j$ represent various multipole deformations of the fuzzy
sphere carrying representation $j$. Similarly, block $b$ represents
multipole deformations of the other sphere $j'$. The blocks $c,
c^\dagger$ represent deformations which involve both spheres and will
be shortly identified with tachyonic directions which deform the spin
$j\oplus j'$ solution to other solutions, e.g. the irrep $j+ j'$.

One can easily check that the results mentioned above apply to all the
tables in the Appendix A.

\subsection{The puzzle}

We find that $\omega^2 \ge 0$ around all the solutions \eq{rep} or
\eq{irrep} that we have studied so far.  This is a puzzle. For one
thing, it appears to contradict the cascade picture mentioned in
Section \ref{cascade}.  Furthermore it is also mathematically absurd
to have a potential none of whose critical points have unstable
eigenvalues.  To put it differently, if all the solutions mentioned in
table 1 are locally stable extrema with no unstable direction then it
points to existence of barriers separating different extrema labelled
by SU(2) representations. If so then there has to be some unstable
extremum separating two such stable solutions. It is clear from above
analysis that if such an extremum exists it is {\em not} a SU(2)
representation.  One of the motivations for this work was to resolve
this puzzle. As we will see in the next section, this puzzle is
resolved in an interesting way. The resolution lies in the fact that
there are new critical points which we will find.

\section{New critical points}

In this section, we will reconsider the equation of motion \eq{eom}
and look for the general static solution and its properties. Notice
that the equations of motion used so far, \eq{comm} and \eq{su2}, for
obtaining the solutions as SU(2) representations are themselves a kind
of ansatz. It is easy to see that solution to these equations solve
the original equation of motion. The converse, however, is not
true. The original equation of motion supports many more solutions
which, in general, are not SU(2) representations. Let us consider the
following situation
\bea
\label{new}
X_i &=&  \bar X_i + c_{ia} Y_a \label{newsol}\\
{}[ \bar  X_i, \bar  X_j  ] &=&
i\ \kappa\ \epsilon_{ijk}\ \bar  X_k 
\nonumber\\
{}[Y_a, Y_b] &=& 0 = [Y_a, \bar X_i].
\label{commat}
\eea
This is clearly a solution of the equation of motion \eq{eom}, whereas
it does not solve \eq{comm} or \eq{su2}.

In the above, $Y_a, a=1,2,\cdots, n_a$ are any set of matrices
satisfying \eq{commat}, and $c_{ia}$ are any $3 n_a$ real
numbers. Since we are concerned here with only traceless $X_i$'s we
will consider only traceless $Y$'s.  These constitute finite
deformations of the solution $\bar X_i$ discussed in the last section.
These finite deformations effected by $Y_a$, by virtue of \eq{commat},
do not change energy of the solution. Energy of this solution is the
same as that obtained using $\bar X_i$ as a solution. The coefficients
$c_{ia}$ give a continuous family of degenerate solutions, which in a
sense parametrise the moduli space of these solutions. Another
way of saying it is that the $Y$-deformations are exactly marginal
or flat directions.

As an example, suppose $\bar X_i$ are given by the solution (3) of
Table 1.  
%The matrix $\bf{1}\otimes \vec{\sigma_3}$ satisfies line (3)
%of the equation \eq{new} 
We get a family of solutions labelled
by $\vec c$.  The new solution (family) \eq{new} is, therefore
\be
X_i = \kappa \left( \begin{array}{cc}
		    \tilde{J}_i  & {\bf 0} \\
		   {\bf 0} &   \tilde{J}_i
  \end{array} \right)
+  c_i\ Y, \; Y= \left( \begin{array}{cc}
		   {\bf 1} & {\bf 0} \\
		   {\bf 0} & -{\bf 1}  \end{array} \right)
\label{half-half-new}
\ee
Here $\tilde J_i \equiv 1/2 \sigma_i$ are the spin 1/2 repn. of SU(2).
In this example, there is only one marginal deformation $Y$.
The specific choice $Y={\bf 1} \otimes \sigma_3$ could be replaced
by $Y={\bf 1} \otimes d_i \sigma_i$ which still 
satisfies \eq{commat}.

Now that we have the new solutions, let us now
go back to the analysis of quadratic fluctuations
\eq{fluct}. As we will see, each of these new solutions represents
a collection of spherical D-brane solutions of various sizes whose
centres are separated by the $Y$-parameters above. The particular case
where the $Y$-deformations refer to locations of D0-branes separated
from a single spherical brane has been discussed in great detail
earlier in \cite{Hashimoto:2001xy}.

\subsection{Spectrum of
fluctuations after $Y$-deformation}\label{perturb}

We now consider the eigenvalues of the quadratic fluctuation operator
in \eq{fluct} after we turn on the $Y$-deformations.  Although the
deformations do not change energy, the quadratic fluctuations around
the deformed solution are rather different from those around the new
solution. In particular, the third term in \eq{fluct} which vanished
by virtue of being proportional to the equation of motion \eq{su2}
does not vanish any more. Earlier results on fluctuations can be
obtained by turning off $Y_a$ deformations. Quadratic fluctuations
around a general solution are calculated in Appendix A.

\subsubsection{Tachyons} 

As we can see, negative eigenvalues of $\delta^2 H$ appear around
these new solutions.  In some cases, the instabilities appear after a
finite amount of marginal deformation; in some other cases, even
infinitesimal deformations lead to solutions with tachyons. Let us
consider the $4\times 4$ matrix problem. In this case as mentioned in
the previous section we have five inequivalent solutions. To show how
tachyonic instability is obtained in different situations we will
concentrate on three cases of spin $1/2 \oplus 0 \oplus 0$, $1/2
\oplus 1/2$ and spin $1\oplus 0$.

In case of single spin 1/2 representation, the background, i.e., the
solution to the equation of motion \eq{eom}, is given by 
\be X_i =
\lambda_i+ c_i\lambda_8+d_i\lambda_{15}, 
\ee 
where, $i=1,2,3$ and
we have used Gellmann's notation for SU(4) generators.  That is,
$\lambda_i, i=1,2,3$ are spin 1/2 generators (Pauli matrices) in the
upper left $2\times 2$ block (and zero in the rest). $\lambda_8$ and
$\lambda_{15}$ are traceless Cartan generators which are proportional
to identity in the upper diagonal $2\times 2$ block. This solution
geometrically represents a $S^2$ corresponding to spin 1/2
representation and two isolated D0 branes located at $\vec c$ and
$\vec d$. In the absence of marginal deformations, i.e., deformations
along the flat directions, multiplicity of massless fluctuations is
23. By turning on infinitesimal marginal deformation, four massless
fluctuations acquire mass. While two of them have positive mass
square, the other two are tachyonic, signifying instability of this
configuration towards formation of more stable ones. These more stable
configurations are obtained by allowing these tachyonic modes to
condense.  Instead of allowing these tachyon modes to condense, if we
deform further along the marginal direction then for the deformation
parameter $(\sqrt{3}-\sqrt{2}) < 2c < (\sqrt{3}+\sqrt{2})$, we see
that there is another tachyonic instability. One of the novel features
of this instability is that due to the finite marginal deformation,
some of the irrelevant operators become relevant or equivalently
tachyonic. Later we will discuss other novel features of these new
unstable directions.

Now let us look at the spin $1/2 \otimes 1/2$ solution. This solution
is represented as \be X_i = \sigma_i\otimes I + c_i I\otimes\sigma_3,
\ee where, $c_i$ denote marginal deformations.  This is the first
instance when one encounters a solution consisting of two nontrivial
SU(2) representations. This solution also develops tachyonic
instability due to infinitesimal marginal perturbation. In this case
also two marginal directions become relevant or tachyonic. Like the
first case here also after a finite marginal deformation there is
another tachyonic instability, where the marginal parameter the
marginal parameter takes values in the interval $1-\sqrt{1/2} < c <
1+\sqrt{1/2}$. The situation is quite different for the solution $1
\otimes 0$. 

The spin $1\otimes 0$ solution is represented as \be X_i = J_i +
c_i\lambda_{15}, \ee where, $J_i$ consist spin 1 generators which
occupy $3\times 3$ upper left diagonal block and zero in the last
diagonal entry. In this case there is no tachyonic instability due to
infinitesimal deformation. Even more interesting feature is that no
marginal deformations become massive. This fact is valid for finite
marginal deformations as well. This configuration is not the lowest
energy configuration and hence it is imperative to find the tachyonic
directions in the moduli space of this solutions. As mentioned a
moment ago, none of the marginal directions become tachyonic instead
we find that irrelevant directions become relevant. Since there is no
tachyonic instability due to infinitesimal marginal deformation and
the first time we encounter the instability when the marginal
deformation parameter $ 2c > \sqrt{6} -\sqrt{3}$, the stability radius
$r$ for this solution in the moduli space is given by $ r = (\sqrt{6}
-\sqrt{3})/2$.

\subsection{Enhanced symmetry}

Here we will comment on yet another novel feature of the fluctuation
spectrum. As mentioned above every solution has a multi-dimensional
moduli space. However, we will see later that some of these moduli are
gauge artifacts. Remaining moduli parameters or zero modes are
genuine, at least at the point in the moduli space where the solution
can be written in terms of direct sum of SU(2) representations. In the
previous subsection we saw that some of these zero modes do get lifted
due to infinitesimal marginal deformations. We also encountered a
situation when an irrelevant deformation became relevant after a
finite amount of marginal deformation. In other words a deformation
with positive mass squared becomes one with negative mass squared. A
new massless mode which separates these two regions at a point in the
moduli space produces a new marginal direction. This sudden jump in
the dimensionality of the moduli space is quite reminiscent of
emergence of new marginal operators in c=1 CFT\cite{Dijkgraaf:1988vp}.
Consider an example of spin $1\otimes 0$ solution to the $4\times 4$
matrix problem. This solution develops instability only when the marginal
deformation parameter $ 2c > \sqrt{6} -\sqrt{3}$. In fact, when
$ 2c = \sqrt{6}-\sqrt{3}$ this solution has two additional marginal
directions.

\subsection{Symmetry breaking patterns}

As indicated above, although the $Y$-deformations do not change
energy, they are not trivial symmetry directions. The table of
eigenvalues in Appendix A follow the symmetry breaking pattern.
Recall, in the absence of RR four form background, trivial SU(2)
representation was the lowest energy (BPS) configuration. When $N$ D0
branes are located at one point in space, they give rise to $U(N)$
local gauge symmetry. If these D0 branes are moved away from each
other, this $U(N)$ symmetry breaks down to $U(1)^N$.

In the presence of RR four form flux, this picture changes
dramatically. Now the trivial SU(2) representation has highest energy
among possible extrema.  Though it is a kind of maxima, there are no
relevant directions when all the D0 branes are coincident. It has a
large number of zero modes as argued in the previous subsection,
though several of them are gauge artifacts. Thus this extremum
resembles, in the language of super Yang-Mills theory, the Coulomb
branch, dimension of which is given by number of nontrivial zero
modes.  However, some of these flat directions are lifted as
infinitesimal deformation along any of the nontrivial marginal
directions reveals instability of this solution to formation of
non-trivial SU(2) representations. Now let us consider one nontrivial
SU(2) representation, say, spin 1/2 and rest all singlets. This
solution has lower energy than the trivial solution and in this case
$U(N)$ gauge symmetry breaks down to $U(N-2)\times
U(1)$\cite{Polchinski:2000uf}. Like the
trivial solution this one also has no relevant operators exactly at
the point where we have SU(2) symmetry. Again we find that
infinitesimal deformation along one of the several marginal (flat)
directions show that there are relevant directions leading to more
stable configurations. Generically, if we have one nontrivial
representation of spin $j$ and rest all are trivial representation
then the gauge group is $U(N-2j-1)\times U(1)$. If instead we have
several nontrivial representations then the residual symmetry is
\[
U(N- \sum_{l=1}^{m} (2j_l+1))\otimes U(1)^m.
\]
The $U(1)$ symmetry associated with each fuzzy sphere gets enhanced if
the reducible solution contains multiple number of SU(2)
representation. For example, in case of $4\times 4$ matrices, we get a
solution which contains 2 SU(2) spin 1/2 representations. Apart from
the U(1) symmetry of these configurations, we also have additional
symmetry which rotates these two spin 1/2 configurations into each
other without any energy cost. This enhances the gauge symmetry to
SU(2). The biggest sphere, which corresponds to $j=(N-1)/2$ spin
solution is the most stable configuration. This solution has no
nontrivial flat directions and in this case the gauge symmetry is
completely Higgsed. Since there are no relevant as well as marginal
directions, the spectrum has a gap and therefore this solution
represents the Higgs branch.

We will now illustrate this by taking an example of $4\times 4$
matrices.  Among the five classical solutions, the trivial solution
preserves $U(4)$ gauge symmetry. The solution with spin $1/2\oplus
0\oplus 0$ has $U(2)\otimes U(1)$ symmetry, where $U(2)$ is the
symmetry of two coincident D0 branes and $U(1)$ is contributed by the
fuzzy sphere associated with spin 1/2 representation. The extremum
corresponding to spin $1\oplus 0$ breaks the symmetry down to
$U(1)\otimes U(1)$. The minimum energy solution corresponds to spin
3/2 and in this case there is no residual gauge symmetry left over in
the problem. As mentioned earlier, $4\times 4$ matrix also has a
solution with spin $1/2\oplus 1/2$. Since the spin 1/2 representation
appears twice in this solution we have an enhanced symmetry, which
rotates two spin 1/2 representations into each other. This results in
an additional U(2) gauge symmetry of the solution.

\section{Comparison with other Approaches}

In this section we will mention other related works and compare the
results obtained here with those obtained in the other approaches. For
facility of comparison, we will reproduce our results from Sections 2
and 3 about the radius $R_N$ and the energy $E_N$ of the irrep and the
eigenvalue $w^2_l$ of the quadratic fluctuation operator $M$
\eq{fluct}:
\bea
R_N &=&  \frac\kappa{2} \sqrt{N^2-1}
\label{our-rn}\\
E_N &=&  - \frac{T_0}{24} \kappa^4 N(N^2-1)  
\label{our-en}\\
w^2_{l,N} &\equiv& \kappa^2 \omega^2_l = \kappa^2 l(l+1),
\quad l=0,1,\ldots,N-1
\label{our-wn}
\eea

\subsection{Spherical D2-branes in $S_3 \times R^7$} 

This model is studied in \cite{Bachas:2000ik}. Strictly speaking, this
is not a valid background of string theory (has dilaton tadpoles etc,
for instance), but for the purposes of the present section this fact
can be ignored.  The string theory is described by a level-$k$ SU(2)
WZW CFT, (the D2-branes are described by boundary states of this
CFT). \cite{Bachas:2000ik} also presents an alternative discussion in
terms of a Born-Infeld-Chern-Simons (BICS) action for D2 brane. 
We will discuss the latter first.

\subsubsection{Born-Infeld (BICS) description}

This description uses the following closed string background
\bea 
ds^2 &=& \frac{k}{2\pi}(d\psi^2 +
\sin^2\psi (d\theta^2+\sin^2\theta d\phi^2)) \nn 
B &=& \frac{k}{2\pi}
(\psi - \frac{\sin 2\psi}{2}) \sin\theta d\theta d\phi 
\nn
F &=&
-\frac{N}{2} \sin \theta d\theta d\phi 
\label{bachas-bgd}
\eea 
The region of validity of the DBI description is \bds:
\be
k \gg 1, N \gg 1, k/N \gg 1
\label{dbi-limit}
\ee
As \bds shows, a spherical D2-brane in
such a background, wrapping
some $S^2 \in S^3$ is energetically stable, provided that
the area of the $S^2$ is chosen as
\be
{\rm Area}= 4\pi \frac{k}{2\pi}\sin^2\psi_N,
\quad
\psi_N = \frac{\pi N}k
\ee
Here $N$ refers to the quantized $F$-flux, or
equivalently, the number of D0-branes.
This corresponds to a radius 
\be 
R_N = \sqrt{\frac{k}{2\pi}} \sin \frac{\pi N}k
= N\sqrt{\frac{\pi}{2k}}\Big(1 + O(\frac{N}{k})^2 \Big)
\label{dbi-rn}
\ee 
In the region \eq{dbi-limit}, this agrees with \eq{our-rn}, 
provided we choose
\be
\kappa^2=\frac{2\pi}{k} = \frac{1}{k\alpha'}
\label{match}
\ee
Note that we are working in the
convention
\be
2\pi \alpha'=1
\label{def-alpha}
\ee
The mass of the D2-brane turns out to be  \bds
\bea
M_N &=& N T_0 + E_N,
\nn
E_N &=& 2kT_2\sin \frac{\pi N}{k} - N T_0 = 
- T_0\frac{N^3\pi^2}{6
k^2}\Big(1 + O(\frac{N}{k})^2 \Big)
\label{dbi-en} 
\eea 
This also agrees with \eq{our-en} under
\eq{match} and \eq{dbi-limit}. Note that
in our convention \eq{def-alpha} 
$T_0 \equiv 4\pi \alpha' T_2= 2 \pi T_2$.

The eigenvalues of
quadratic fluctuation operator  are 
\be 
w^2_{l,N}= \frac{l(l+1)}{k\alpha'}, l=0,1,\ldots, N-1
\label{dbi-wn}
\ee
which agrees with \eq{our-wn}.

\subsubsection{BCFT description}

The BCFT results for the energy and the
quadratic fluctuation eigenvalues  are \bds
\be
E_N= 2(k+2) T_2 \sin(\frac{N\pi}{k+2})
\ee
\be
w^2_{l,N}= \frac{l(l+1)}{\alpha'(k+2)}.
\ee
It has already been shown in \bds that these
results agree with the Born-Infeld results
in the region \eq{dbi-limit}.

\subsubsection{The ARS matrix model}

The ARS matrix model \ars\ is related to the
formulation of the above system as
a bound state of D0-branes. The
matrix model is given by
\be
S= T_0 Tr [ \frac{2}{k} \dot \Phi_i^2 + \frac{1}{k^2} 
[\Phi_i,\Phi_j]^2
-\frac{4 \sqrt{\pi}i}{3k^2}\epsilon^{ijk} \Phi_i \Phi_j \Phi_k)]
\ee
This matrix model itself agrees with the matrix model
that we have presented \eq{action}, provided
relate $\Phi_i$ to the $X_i$ as follows
\be
\Phi_i = \sqrt{\frac{k}{2}} X_i
\qquad 
\kappa=\sqrt{\frac{2\pi}{k}}
\label{ident}
\ee
The results for the radii, energy and the
fluctuation spectrum of course should agree since
the matrix models themselves agree. The relation 
between $k$ and $\kappa$ is the same as in \eq{match}.

\subsection{D2-branes in IIA in the presence of RR flux}

This is the situation considered in \myers.  The matrix model
(D0-description) is identical to the one in \eq{action}.  The dual
formulation in terms of the world-volume theory of the D2-brane uses
BICS action in the presence of a RR-flux background
\be
F_{tijk}= -2f \epsilon_{ijk}
\ee
and a U(1) flux on the world-volume
\be
F = \frac{N}{2} \sin \theta d\theta \wedge d\phi
\ee
We should identify
\be
f= \kappa
\label{f-kappa}
\ee
The BICS action is given by 
\be
S= 4\pi T_2 \sqrt{R^4+\frac{N^2}{4}}- \frac{8\pi T_2 R^3}{3}
\ee
As discussed in Myers, this action has extrema when 
\be
R= \frac{N f}{2} = \frac{N \kappa}{2}
\label{myers-rn}
\ee 
At these values of the radius of the 
D2-spheres, the energy evaluates to be 
\be
E_N =-\frac{T_0}{24}N^3 f^4
\ee
Under the identification
\eq{f-kappa}, we get perfect agreement with the results of our
matrix model to leading order in $1/N$.

Regarding the fluctuation spectrum, it is easy
to work out the quadratic fluctuations of the BICS
action around the spherical solution for the
D2-brane. The calculation is similar to the one
presented in \bds. The result is
\bea
w^2_{l,N} &=& 
\frac{l(l+1)}{R'}
\nn 
R' &=& \frac{\sqrt{(R^4 + (N/2)^2)}}{R} \approx \frac{N}{2R}^2
\label{myers-wn}
\eea
Thus, the fluctuation spectrum is of the form of a Kaluza-Klein
spectrum on a sphere of effective radius $R'$.  Because of the
relation between $R'$ and $R$, the KK spectrum goes as $R$ in stead of
$1/R$.  The fact that the KK spectrum can be modified because of
noncommutativity has been found elsewhere, e.g.  in
\cite{Das:2001fu,Das:2001st,Gomis:2000sw}.
Because of \eq{myers-rn}, the above fluctuation spectrum
\eq{myers-wn} exactly agrees with
our result \eq{our-wn}.

\section{Dynamics}

In the previous sections we developed a fair idea of the potential
energy landscape of our model.  We found in Section 3 that the cascade
picture as suggested in Section \ref{cascade} is not quite valid;
however there are exactly marginal deformations of the reducible
representations (Section 4). In terms of the energy landscape these
flat directions are like ridges; we found in Section 4 that if we
start from any critical point corresponding to a reducible
representation, and move along the ridges, then downward slopes
develop along which one can roll down to lower energy critical points.
Clearly such a landscape offers rich dynamics. In the present paper we
will not attempt to solve the full dynamics of the $X^i(t)$ which is a
formidable problem; we will instead focus on small submanifolds
of the configuration space which captures important features
of the above landscape and discuss the dynamics within such
ansatzes.

To be concrete, let us consider dynamical evolution from one
SU(2)-representation $\tilde J_i$  another, say $J_i$.
We will discuss this dynamics within the following
ansatzes in turn.

\subsection{One-parameter ansatz} 
 
We wish to solve the \eq{eom}
with the  boundary condition
\be
X_i(t_0) = \kappa \tilde J_i
\label{bc-orig}
\ee
Let us try the following ansatz
({\it cf.} Bachas-Hoppe-Pioline\cite{Bachas:2001dx}) 
\be
X_i(t) = \kappa \left(\tilde J_i + f(t)(J_i - \tilde J_i) \right)
\label{ansatz}
\ee
where $J_i, \tilde J_i$ are two SU(2) representations which we wish to
connect by a dynamical trajectory parametrized by $f(t)$,
\eq{bc-orig} translating to
\be
f(t_0)=0, f(t_1)= 1.
\label{bc}
\ee
We note that for $\tilde J_i=0$, the above ansatz is similar to the
spherical ansatz of the giant graviton scenario, where the energy is
viewed simply as a function of a single parameter of the (D0 or D2)
configuration, namely the radius. Since the flat
directions and the resulting new solutions of Section 4 represent
non-spherical deformations, we
will need at least a
two-parameter ansatz to reproduce the physics of the
flat directions. 

To begin with, however, we will stick with the simple ansatz
\eq{ansatz}.
We discuss the case where $J_i$ is the $N= 2j+1$-dimensional irrep, with $N$
even, and $\tilde J_i$ corresponds to spin $j'\oplus j',
j'=(j-1/2)/2,\hbox{i.e.}  N'\equiv 2j'+1= N/2$. 

Using \eq{ansatz} in \eq{eom}
leads to
\be
\ddot f = - V_j'(f), V_j(x)= -a_j + b_j x^2 - c_j x^3 + d_j x^4
\label{aeom} %ansatz-eom
\ee
where the coefficients depend on the
spin $j$. $a_j, b_j, c_j$ and $d_j$ are found to be positive.

The above problem \eq{aeom} is that of an asymmetric double well
potential (see Figure \eq{a-well}) if we consider $f(t)$ as the
position of a particle on a line.

%\begin{figure}[ht]
\FIGURE{   \vspace{0.5cm}
\centerline{
   {\epsfxsize=5.5cm
   \epsfysize=5cm
   \epsffile{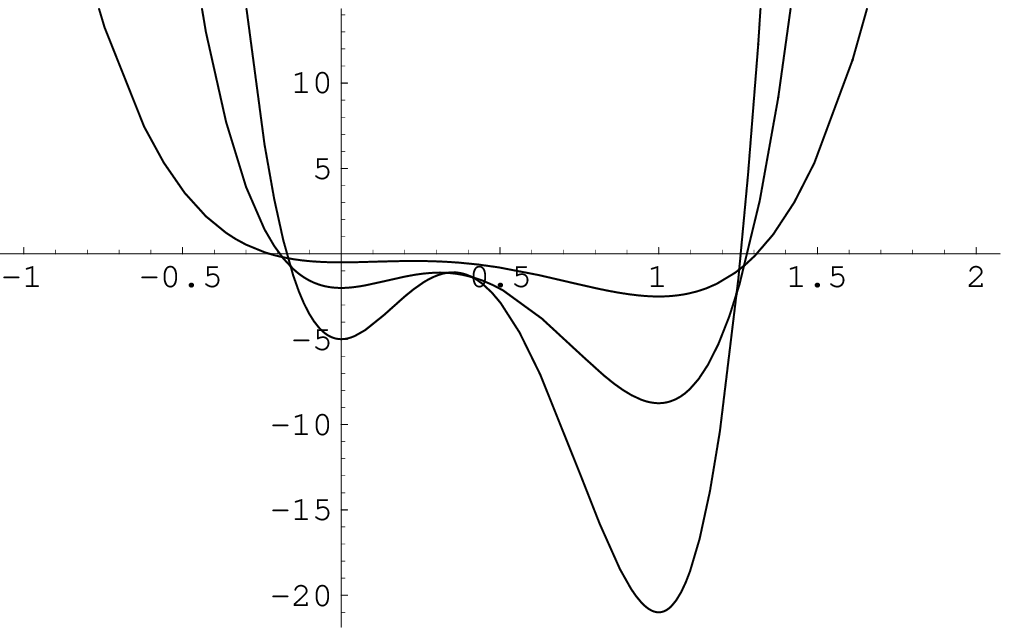}}
}
\caption{\sl The asymmetric double-well for $j=3/2,5/2,7/2$}
\label{a-well}}
%\end{figure}

In Fig. \eq{a-well}, the representation $\tilde J_i$, corresponding to
the double-sphere, is the local minimum on the left whereas the
irreducible representation $J_i$ corresponds to the absolute minimum.
Note that within this ansatz it is not possible to {\em classically}
evolve from $\tilde J_i$ to $J_i$. This is of course similar to the
puzzle referred to in Section 3. Just as that puzzle
was solved by the discovery of flat directions, we need here
to include one more parameter in our ansatz.

\subsection{Two-parameter ansatz including marginal deformations}

The asymmetric double-well potential,
arising from the ansatz \eq{ansatz} misses the flat
direction present in the actual problem. Let us try
a new ansatz which has a new
term involving the flat direction $Y$ (cf. \eq{half-half-new})
\be
X_i(t) = \kappa \Big(
\tilde J_i + f(t)(J_i - \tilde J_i) + g(t) d_i Y \Big)
\label{new-ansatz}
\ee
subject to the boundary condition
\bea
f(t_0) &=& g(t_0)=0
\nonumber\\ 
f(t_1) &=& 1, g(t_1)=0
\label{new-bc}
\eea
The action \eq{action} evaluated on
the trajectory \eq{new-ansatz} for $j=3/2$ i.e $4\times 4$ matrices,
becomes $S= \int dt(K - V)$ with
\be
K=2\left((4-\sqrt{3}){\dot{f}}^2 + 2\dot{f}\dot{g}  + \dot{g}^2\right)
\ee
We can diagonalise the kinetic term by using $g=h-f$. Then the potential term  is
\be
V= -\frac{1}{2}+ 4\left( 2(h-1)^2+(1-\sqrt{3})\right) f^2 - 8 (2-\sqrt{3}) f^3 +
(10-4\sqrt{3}) f^4
\ee
The plot of $V(f,g)$
(Figs \eq{f-g} and \eq{f-h-contour})
shows an energy landscape for a particle
moving in two dimensions (coordinates $f(t), g(t)$).

%\begin{figure}[ht]
\FIGURE{\vspace{0.5cm}
\centerline{
   {\epsfxsize=5.5cm
   \epsfysize=5cm
   \epsffile{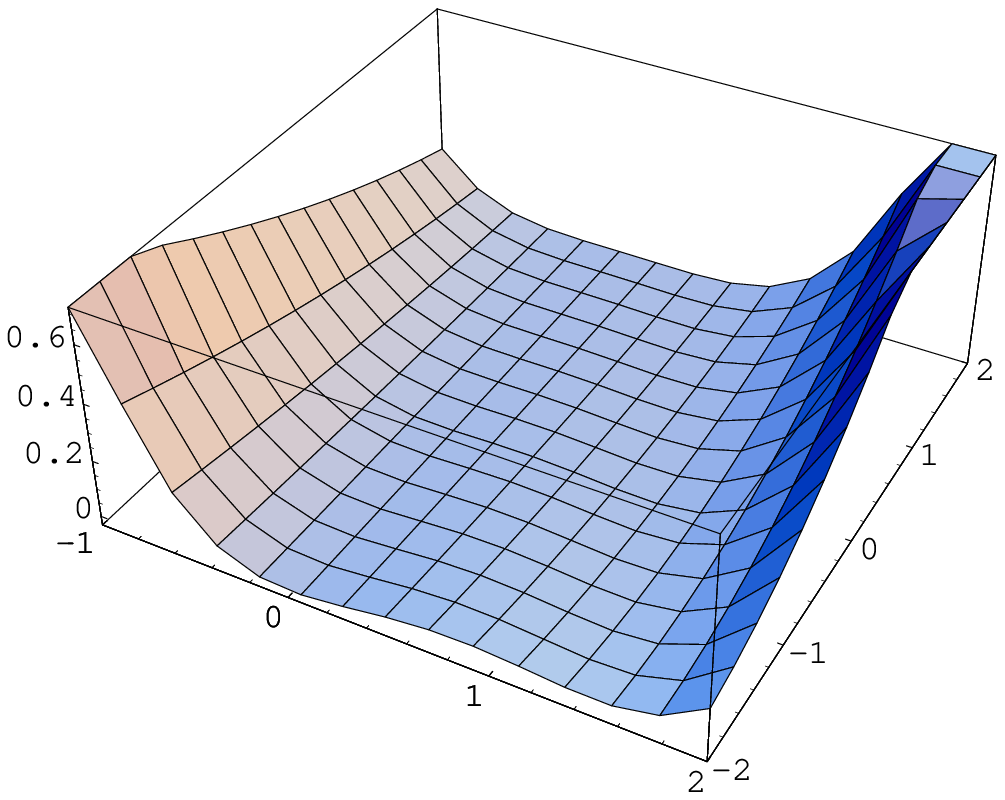}}
}
\caption{\sl The potential $V(f,g)$  for $j=3/2$.}
\label{f-g} }
%\end{figure}
%\begin{figure}[ht]
\FIGURE{\vspace{0.5cm}
\centerline{
   {\epsfxsize=5.5cm
   \epsfysize=5cm
   \epsffile{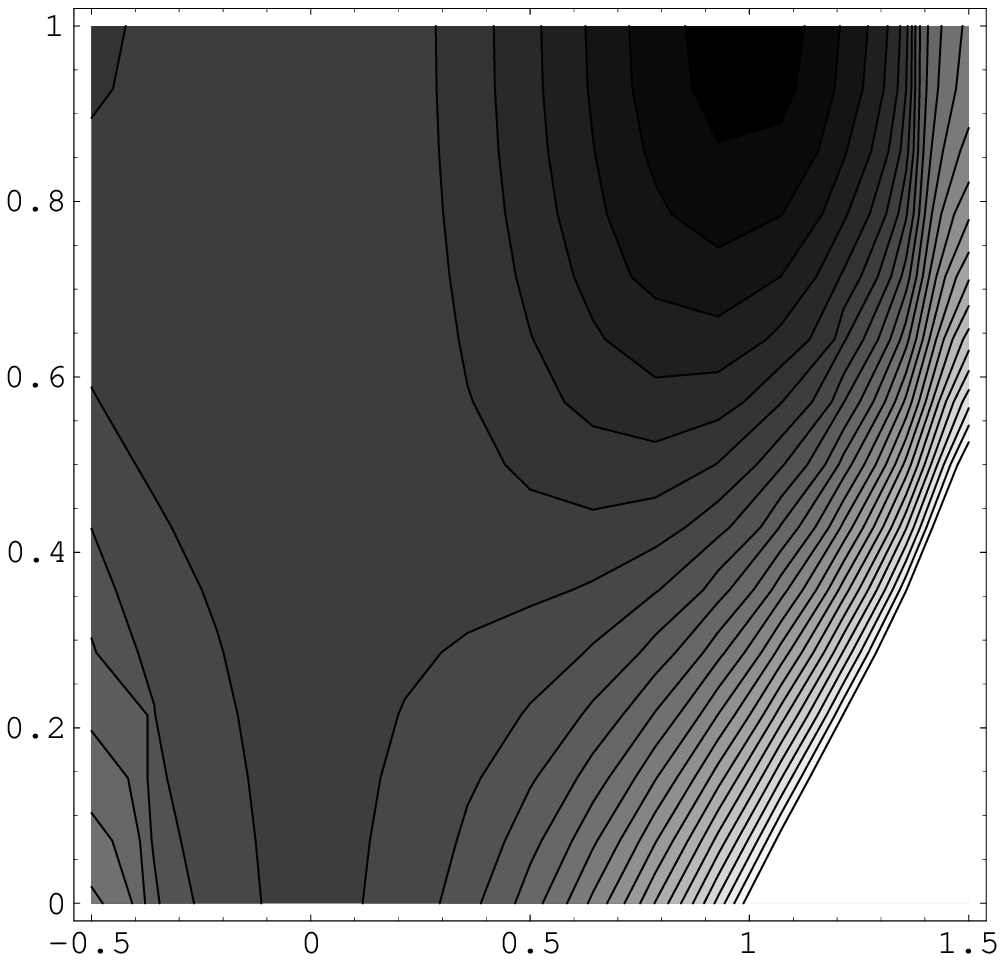}}
}
\caption{\sl Contour plot of $V(f,h)$  for $j=3/2$.}
\label{f-h-contour}}
%\end{figure}

It is clear from Fig \eq{f-h-contour} that it is possible to roll down
reducible representations $\tilde J_i$ (the point $f=h=0$) to the
irreducible representation $J_i$ (the point $f=h=1$) 
without experiencing an energy barrier.

In \cite{MWY}, we develop a geometrical picutre of this dynamical
process, where two spheres merge into one through non-spherical
deformations.

To conclude this section, we have shown that within the
ansatz \eq{new-ansatz} it is possible to roll down
from a reducible representation to an irreducilbe representation.

\section{Effect of adding a mass term}

In this section we present briefly the effect of adding
mass terms to our model \eq{action}:
\be
S= T_0\int dt {\rm Tr}({1\over 2}\dot{X}_i^2 +
\frac{1}{4}[X_i,X_j]^2-\frac{2i\kappa}{3}\epsilon_{ijk}X^i X^j X^k-
\frac{m^2}{2} X_i X_i)
\label{action-m}
\ee
The static equation of motion is
\be
[X_j, \bigg( [X_i,X_j] - i\kappa\  \epsilon_{ijk} X_k \bigg) ]-m^2 X_i=0.
\label{eom-m}
\ee
Once again this admits solutions of the
form \eq{rep}
\be
X_i=a \oplus_{r=1}^s J_i^{(r)}
\label{rep-m}
\ee
where now $m^2=2(a\kappa-a^2)$, i.e.
\be
a=(\kappa \pm \sqrt{\kappa^2-2m^2})/2
\ee
The lower sign gives an unstable solution, so
we shall only consider the upper sign henceforth.

The model is supersymmetric for $m= \frac{2\kappa}{3}$
\cite{Polchinski:2000uf}.  However, for the time being we will work with
generic $m$. In the table below we list the quadratic fluctuation
spectrum and the value of the action around \eq{rep-m}.

\vspace{5ex}

\vbox{
\begin{center}
Table 3: Fluctuation spectrum with mass term\\

\begin{tabular}{||l|l|l|l||}\hline
 Solution & $\omega^2$ & Multiplicity& Action\\ \hline
  	    & 0 	&   7&  \\ \cline{2-3}
&  $m^2$		  & 3&  \\ \cline{2-3}
$\frac{1}{2}\oplus 0$	&$3(\kappa^2+3m^2+\kappa\sqrt{\kappa^2-2m^2})/8$ & 8
& $\frac{a^2}{4}(3a^2-4a\kappa+3m^2)$ \\ \cline{2-3}
	&  $(\kappa^2+m^2+\kappa\sqrt{\kappa^2-2m^2})$  & 5 & \\ \cline{2-3}
        &  $\kappa^2-2m^2+\kappa\sqrt{\kappa^2-2m^2}$   & 1 & \\ \cline{1-4}
             & 0 & 12&  \\ \cline{2-3}
 	$\frac{1}{2}\oplus \frac{1}{2}$& $m^2$& 9&
  $\frac{a^2}{2}(3a^2-4a\kappa+3m^2)$   \\ \cline{2-3}
&  $(\kappa^2+m^2+\kappa\sqrt{\kappa^2-2m^2})$& 20 & \\ \cline{2-3}
& $\kappa^2-2m^2+\kappa\sqrt{\kappa^2-2m^2}$  & 4   &\\ \cline{1-4}
\end{tabular}\\
\end{center}
}
\vspace{0.2cm}

It is straightforward to see that we get the original
energy spectrum in absence of $m$. On the other hand, when $m=2\kappa/3$
all the solutions have zero energy which is consistent with their being
supersymmetric vacua.

% {}From $m=0$, which is our starting point to $m=2\kappa/3$,
% which is the Polchinski Strassler point,
% the $a^2-m^2$ eigenvalue decreases from $\kappa^2$ to zero.
% So there is no tachyon in this regime ( $\kappa>0$ assumed).

Let us rescale the field $X_i/\sqrt{2} = Y_i$ and substitute it in the
energy functional of this modified matrix model
\be
E = {\rm Tr} \int dt \Big({\dot Y_i}^2 - [Y_i,Y_j][Y_i,Y_j] + m^2 Y_iY_i
+{2\sqrt{2}i\kappa\over 3} \epsilon_{ijk}Y_i[Y_j,Y_k]\big).\label{yaction}
\ee
We can write this in terms of a total square term, a total derivative term and
a defect term.
\bea
E &=& {\rm Tr} \int dt \left[\dot Y_i \pm \left({2\sqrt{2}\over 3}\kappa
Y_i + {i\over 2}\epsilon_{ijk}[Y_j,Y_k]\right)\right]^2
+ {\rm Tr} \int dt (m^2-{4\kappa^2\over 9})Y_i^2\nn
&\mp& {\rm Tr} \int dt {d\over dt} \left( \left({2\sqrt{2}\over
3}\kappa\right) Y_i^2 + {i\over
3}\epsilon_{ijk}Y_i[Y_j,Y_k]\right).\label{energy} \eea
Lowest energy configurations are those which equate the first term to
zero. This gives us the first order equation. Second term in the energy
functional is the defect term. As it stands it is also a total square term but
equating it to zero gives us only trivial solution. It is this term which
vanishes in the supersymmetric limit, giving rise to the classical form
of the Bogomolnyi energy functional. In the Bogomolnyi limit, energy
functional factorises into perfect square terms and total derivative
terms. While setting former to zero gives us first order equations of
motion, latter gives us the energy/mass of the solution. For the time
being we will treat the second term in the energy functional as a defect
term and will not, either set it to zero or demand it as a constraint on
the solution. The first order equations of motion are
\be
\dot Y_i \pm \left({2\sqrt{2}\over 3}\kappa Y_i +
{i\over 2}\epsilon_{ijk}[Y_j,Y_k]\right) = 0.
\ee
This equation has been studied by Bachas, Hoppe and
Pioline\cite{Bachas:2001dx}. A class of solutions studied by them are those,
which interpolate between trivial SU(2) representation and, say, the largest
possible irreducible representation of SU(2). The Bachas, Hoppe and Pioline
ansatz which solves the equation of motion is
\be
Y_i = {\sqrt{2}\kappa/3\over {1 + \exp{(\sqrt{2}\kappa(t-t_0)/3)}}}\rho_i
\ee
where $\rho_i$ is some SU(2) representation. Contribution of the total
derivative term to the energy of this configuration comes only from
the endpoints of the trajectory, i.e., from difference of the energy
of spin 0 representation and that of representation $\rho_i$. In
addition there is a contribution from the defect term. For the
solution under consideration, this defect term contributes infinite
energy. Recall that in the supersymmetric limit this defect term
vanishes and hence there is no infinite contribution to the energy of
this solution. Thus this is a finite energy solution in the
supersymmetric problem. Of course, in the supersymmetric case, energy
of every vacuum configuration is zero, and hence this solution
interpolates between two degenrate vacua. On the contrary, in the
non-supersymmetric case, vacuum configurations have different energy
which depends upon their SU(2) labels. It is this non-degeneracy of
these vacua which is responsible for the infinite energy of these
solutions. For large enough $N$, even in the non-supersymmetric
scenario, we have degenerate vacua which are labelled by different set
of SU(2) indices. If we consider a solution interpolating between
these vacua, it will clearly have finite energy.

One of the important implications of the mass term is that the moduli
are lifted. This is evident from the first order equations. Static
solutions to these equations do not allow the marginal deformations
that we studied in section 4. Thus, it is not classically possible any
more to roll down from a reducible representation to an irreducible
one. Only way we can reach the irreducible representation is by
tunneling. The solution discussed above falls in this class.

\section{Comments on SUGRA duals}

\subsection{Polchinski-Strassler scenario}

The matrix model \eq{action-m} coincides with the space-independent
part of the action for the adjoint scalars of the four-dimensional
gauge theory considered in \cite{Polchinski:2000uf}. The
supersymmetric point $m=2\kappa/3$ and its SUGRA dual is discussed
there in great detail. For other values of $m$, supersymmetry is
broken; SUGRA dual of these theories also are briefly mentioned in
\cite{Polchinski:2000uf}. Essentially the SUGRA duals of the classical
solutions \eq{rep-m} of the gauge theory are a collection of 5-branes
in a space which is asymptotically $AdS_5\times S^5$. Close to
themselves, the 5-brane world-volume is like $R^3 \times S^2$.
The radii of the 2-spheres and their location in the AdS geometry
depends on the dimensionality of the representation $J_i^{(r)}$
in \eq{rep-m}.

\cite{Polchinski:2000uf} has indicated a phase diagram in the $(m^2,
\kappa)$ plane where the $m^2<0$ represents a phase in which only the
irreducible representations are stable. The phase $m^2>0$, on the
other hand, is qualitatively similar to the supersymmetric point
$m=\frac{2\kappa}3$; in this phase the reducible representations are
all stable. The matrix model \eq{action} which we began with
has $m^2=0$; as we found in our stability analysis, here
the reducible representations are marginally stable, with
classical instabilities in the cubic order. In terms
of the SUGRA dual, this implies that the collection of
5-branes corresponding to the reducible representations
should have a mild classical instability towards forming
a large 5-brane corresponding to the irredicuble
representation.

\subsection{F1-NS5 brane system}

Another place where this matrix model is relevant is in the F1-NS5
brane system in type IIA string theory. Easiest way to see this is to
start with the D1-D5 system of type IIB string theory. The Chern-Simon
term on the common worldvolume of the D1-D5 system contains a term
($a,b=0,1$ are common world volume indices, $i,j,k,...=2,3,4,5$ are
five-brane directions, transverse directions ignored),
\be
S_{cs,D1D5}
= \int d^2\sigma [ \cdots + i {\lambda\over
2}(C^{(4)}_{abij}[\Phi^j,\Phi^i]+C^{(4)}_{aijk}D_b\Phi^i[\Phi^k,\Phi^j])
+ \cdots]\label{dcsterms}
\ee
where we have ignored terms involving
higher form potentials.  Since the potential $C^{(4)}$ is invariant
under S-duality transformation, identical term exists on the common
world volume of F1-NS5 brane system in type IIB string theory.
\be
S_{cs,NS} = \int d^2\sigma [\cdots + i {\lambda\over
2}(C^{(4)}_{abij}[\Phi^j,\Phi^i]+C^{(4)}_{aijk}D_b\Phi^i[\Phi^k,\Phi^j])
+ \cdots]\label{csterms}
\ee
Relevant type IIA configuration of F1-NS5
branes is obtained by performing T-duality along the F1 string
direction. Suppose F1 string is along $x^1$ then, T-duality along
$x^1$, leads to a term with $C^{(3)}_{tij}$, which comes from
$C^{(4)}_{abij}$ and a term with $C^{(3)}_{ijk}$, which comes from
$C^{(4)}_{aijk}$. Incorporating these effects of T-duality into
\eq{csterms}, we see that the common worldvolume theory of F1-NS5
system in type IIA string theory contains the matrix model under
consideration. The commutator square term comes from the DBI part of
the action. To be able to do T-duality along the F1 direction,
i.e. $x^1$, we require all the fields (NS-NS as well as R-R) to be
independent of $x^1$.  Therefore, effective common world volume
dynamics essentially reduces down to time evolution only. Hence, this
situation is identical to the D0 brane system.

\section{Summary}

Matrix model in the RR four form background has fuzzy spheres as
classical solutions demonstrated by Myers\myers . In this paper we
studied the energy landscape of these calssical vacua. Contrary to our
expectation we found that the reducible SU(2) representations,
corresponding to multiple fuzzy sphere solutions are all stable. That
is, quadratic fluctuations around these solutions do not have
tachyonic instability although these solutions all have higher energy
the the irreducible representation.  We resolve this puzzle by showing
that the equations of motion allows many more solutions. These new
solutions are energetically degenerate with the known solutions,{\em
viz.} SU(2) representations. In other words, original solutions have a
large moduli space. Quadratic fluctuations around the new deformed
solutions do have tachyonic instability indicating the roll down path
towards more stable configurations. The irreducible representation of
SU(2) has no nontrivial flat direction as well as the quadratic
fluctuation does not have any tachyonic modes. This feature is
expected as the irreducible representation is the lowest energy
solution.  We then compared our results with other approaches,
particularly the spherical D2 brane of Bachas, Douglas and
Schweigert\cite{Bachas:2000ik}, WZW matrix model of Alekseev,
Recknagel and Schomerus\ars, and Myers dual D2 brane in the presence
of RR four form flux.

We then discussed the effect of marginal deformation and studied the
dynamical trajectory which showed that it is classically possible to
roll down from a configuration corresponding to a reducible
representation to an irreducible representation. We also studied the
effect of switching on the mass term in the matrix model. For a
specific value of this mass perturbation, the matrix model becomes
supersymmetric. At this point all the classical vacua become
degenerate with zero energy. Precisely at this point the second order
differential equation of motion can be factored into a first order
equation. The main new observation in this case is that the moduli are
lifted due to mass perturbation. Thus, it is not classically possible
any more to roll down from a reducible representation to an
irreducible one. However, time dependent solutions to the equation of
motion do exist. These tunneling solutions non-perturbatively
interpolate between different vacua.

There are several leads which are worth pursuing. Firstly, full
time-development of the roll down solution in the massless case is
important particularly from the viewpoint of cosmology. Along the same
lines it will be interesting to develop the time-dependent solution in
supergravity (cf. Sec. 8). Secondly, there is an interesting issue
related to topology change due to non-commutativity. In the massless
case, there is a topology change classically\cite{MWY}. On the other
hand, in the massive case, there is no classical path interpolating
between different classical vacua and therefore, there is no topology
change classically, but it can occur via quantum tunnelling
\cite{Grosse:2001ss}.
\vspace{5ex}

\noindent {\bf Acknowledgement}: We would like to acknowledge 
discussions with Sumit Das, Shiraz Minwalla and Sandip Trivedi.

\appendix
\twocolumn
\section{Eigenvalues}
In this appendix, we calculate the eigenvalues $\kappa^2 \omega^2$ of
the quadratic fluctuation matrix $M$ in \eq{fluct} around the general,
static classical solution \eq{new}. We present the calculations for $n
\times n$ matrices for $n=2,3,4,5$.

\begin{flushleft}
\small{
\begin{tabular}{|l|l|l|}\hline
\multicolumn{3}{|c|}{\bf N=2} \\ \hline
Solution    & $\omega^2$       & Multiplicity    \\ \hline
$0 \oplus 0$ & 0                & 5       \\\cline{2-3}
             & $4(\pm c + c^2)$ & 2(each) \\\cline{1-3}
$1/2$        & 0                & 3       \\ \cline{2-3}
             & 2                & 6   \\ \hline
\end{tabular}}
\end{flushleft}
\noindent Here $\vec{c}$ is the separation between the zero branes, i.e the
solution is $X_i= c_i \sigma_3$.

\begin{flushleft}
\small{
\begin{tabular}{|l|l|l|}\hline
\multicolumn{3}{|c|}{\bf N=3} \\ \hline
Solution   & $\omega^2$ & Multiplicity \\ \hline
        &0 & 12 \\ \cline{2-3}
               & $4(\pm c + c^2)$ & 2 (each) \\ \cline{2-3}
$0 \oplus 0 \oplus 0$&$|\vec{c}-\vec{d}|^2 \pm 2|\vec{c}-\vec{d}|$ 
& 2 (each)\\ \cline{2-3}
               &$|\vec{c}+\vec{d}|^2 \pm 2|\vec{c}+\vec{d}|$
& 2 (each)\\ \cline{1-3}
   &        0 & 10  \\ \cline{2-3}
     $\frac{1}{2} \oplus 0$          & 2     & 6     \\ \cline{2-3}
               &$\frac{3}{4}(1 + 4|\vec{c}|^2 \pm 4 \sqrt{3}|\vec{c}|)$     & 2 (each)
\\ \cline{2-3}
               & $\frac{1}{4}(3 + 12|\vec{c}|^2 \pm 4\sqrt{3}|\vec{c}|)$& 2 (each) \\
\cline{1-3}
        & 0    & 8      \\ \cline{2-3}
         $1$        & 2     & 6      \\ \cline{2-3}
               & 6     & 10    \\ \hline
\end{tabular}}
\end{flushleft}

\noindent The solution in the 3 0-branes case, is defined as follows: 
 
$X_i=c_i \lambda_3+ \frac{d_i}{\sqrt{3}} \lambda_8$, 

\noindent where the $\lambda_{3,8}$'s are the Cartan elements. 

\noindent In the spin $1/2 + 0$-brane case, the solution is 

$X_i=\lambda_i+c_i\lambda_8$, 

\noindent where the $\lambda_i$ are non-zero only in the upper left $2\times 2$
block, and form an su(2) sub-algebra.

\begin{flushright}
\small{
\begin{tabular}{|l|l|l|}\hline
\multicolumn{3}{|c|}{\bf N=4} \\ \hline
Solution & $\omega^2$ & Multiplicity \\ \hline
    &0	&19  \\ \cline{2-3}
          &2      & 6 \\ \cline{2-3}
$\frac{1}{2}\oplus 0\oplus 0 $&$\frac{3}{4}(1\pm 4\sqrt{3}c+4c^2)$&2 (each)\\ \cline{2-3}
               &$\frac{1}{4}(3\pm 4\sqrt{3}c+12c^2)$&2 (each) \\ \cline{2-3}
               & $\frac{4}{3}(\|\vec{\alpha}\|^2\pm \sqrt{3}\|\vec{\alpha}\|)$ & 2
(each) \\ \cline{2-3}
               & $e_2$ & 2 (each) \\ \cline{1-3}
      & 0       & 17    \\ \cline{2-3}
               &       2       & 12    \\\cline{2-3}
      $\frac{1}{2} \oplus \frac{1}{2} $          & $4(\pm c+c^2)$ & 2 (each)\\ \cline{2-3}
               & $2(1\pm 4c+2c^2)$& 2 (each) \\ \cline{2-3}
               & $2(1\pm 2c+2c^2)$& 2 (each)            \\ \cline{2-3}
               & $2(1+2c^2)$   &   4  \\ \cline{1-3}
     & 0 & 17  \\ \cline{2-3}
                & 2     & 6  \\ \cline{2-3}
               & 6     & 10     \\ \cline{2-3}
       $1 \oplus 0$&$\frac{2}{3}(3 \pm 2\sqrt{6}c+4c^2)$& 2 (each)  \\ \cline{2-3}
               &$\frac{2}{3}(3 \pm 4\sqrt{6}c+4c^2)$& 2 (each) \\ \cline{2-3}
               &$\frac{2}{3}(3+4c^2)$ & 4  \\ \cline{1-3}
    		& 0 & 15 \\ \cline{2-3}
$\frac{3}{2}$ 	& 2 &  6 \\ \cline{2-3}
    		& 6 & 10 \\ \cline{2-3}
		&12 & 14 \\ \hline
\end{tabular}}
\end{flushright}
$\|\vec{\alpha}\|^2 =
 |\vec{c}|^2+2 |\vec{d}|^2 -2 \sqrt{2}\vec{c}\cdot \vec{d}\equiv
 c^2+d^2-2 \sqrt{2}cd$.
 
$e_2= (1/12)(9 + 4(c+2\sqrt{2}d)^2\pm4\sqrt{3} (c+2\sqrt{2}d)
 \pm 8\sqrt{3}(c+2\sqrt{2}d))$.

\noindent The solution in the spin $1/2 + 2$ 0-branes case, is defined
 as follows: 
 $X_{i}=\lambda_i+\vec{c}\lambda_{8}+\vec{d}\lambda_{15}$,
 
\noindent where $\lambda_i$ are the spin-$1/2$ rotation generators in the upper
 left $2\times 2$ block (and zero in the rest) and $\lambda_{8}$ and
 $\lambda_{15}$ are the Cartan generators which are equal to identity
 in the upper left $2\times 2$ block. This situation represents an
 $S^2$ brane at the origin and two D0-branes located at positions
 labelled by $\vec{c}$ and $\vec{d}$.

\noindent In the spin $1/2 + 1/2$ case, the two spin half 2-branes are separated
 along their center of mass by the vector $\vec{c}$, i.e the solution
 is

 $X_{i}=\sigma_i\otimes I +c_i I\otimes\sigma_3$

\noindent Similarly, in the spin $1 + 0-$brane, the solution is 

 $X_{i}=J_i+c_i \lambda_{15}$ 

\onecolumn

\noindent where $J_i$ are the spin-$1$ rotation generators in the
 upper left $3\times 3$ block, and $\lambda_{15}$ is the
 Cartan element $(1/\sqrt{6})Diag(1,1,1,-3)$.

\begin{center}
\small{
\begin{tabular}{|l|l|l|}\hline
\multicolumn{3}{|c|}{\bf N=5}\\ \hline
Solution & $\omega^2$& Multiplicity \\ \hline
$\frac{1}{2}\oplus 0\oplus $  & 0     & 42 \\ \cline{2-3}
$ 0 \oplus 0$           & $3/4$ & 24 \\ \cline{2-3}
                        & 2     & 6  \\ \cline{1-3}
                        & 0     &28 \\ \cline{2-3}
                        & 2     &6  \\\cline{2-3}
                        & 6     &10  \\ \cline{2-3}
                        &$(2/3)(3+4c^2)$&4  \\ \cline{2-3}
                        &$(2/3)(3\pm 2\sqrt{6} c +4c^2)$& 2 (each) \\ \cline{2-3}
$1\oplus 0 \oplus 0$    &$(2/3)(3\pm 4\sqrt{6}c + 4c^2)$& 2 (each)  \\ \cline{2-3}
                        &$(1/6)(12+|\vec{\beta}|^2)$& 4   \\ \cline{2-3}
                        &$(1/6)(3|\vec{\gamma}|^2\pm
2\sqrt{6\|\vec{\gamma}\|^2})$&2 (each)  \\ \cline{2-3}
                        &$(1/6)(12\pm 2\sqrt{6}|\vec{\beta}|+|\vec{\beta}|^2)$&2
(each)  \\ \cline{2-3}
                        &$(1/6)(12\pm 4\sqrt{6}|\vec{\beta}|+|\vec{\beta}|^2)$&2
(each)  \\ \hline
                              & 0     &32\\ \cline{2-3}
$\frac{1}{2} \oplus\frac{1}{2}\oplus 0$     & $3/4$ &16 \\ \cline{2-3}
    			      & 2     &24 \\ \cline{1-3}
                        & 0 & 26 \\  \cline{2-3}
			& 3/4 & 8 \\  \cline{2-3}
$\frac{1}{2} \oplus 1$		& 2 & 12 \\  \cline{2-3}
			& 15/4 &16 \\ \cline{2-3}
			& 6 & 10 \\ \hline
\end{tabular}}
\end{center}

$\|\vec{\gamma}\|^2=\|\vec{c}\|^2+(5/3)\|\vec{d}\|^2-2\sqrt{\frac{5}{3}}
\vec{c}\cdot\vec{d} $

$\|\vec{\beta}\|^2=\|\vec{c}\|^2+15\|\vec{d}\|^2+2 \sqrt{15}\vec{c}\cdot\vec{d}$

and $c=\|\vec{c}\|$.

\noindent Note that the solution in the spin $1 + 2$ 0-branes case, is defined
as follows: $X_{i}=\lambda_i+\vec{c}\lambda_{21}+\vec{d}\lambda_{24}$,
where $\lambda_i$ are the spin one rotation generators in the upper
left 3X3 block (and zero in the rest) and $\lambda_{21}$ and
$\lambda_{24}$ are the Cartan generators which are equal to identity
in the upper left 3X3 block. This situation represents an $S^2$ brane
at the origin and two D0-branes located at positions labelled by
$\vec{c}$ and $\vec{d}$.

\end{document}